# Student Reasoning About the Divergence of a Vector Field

Charles Baily and Cecilia Astolfi

*School of Physics and Astronomy, University of St Andrews, St Andrews, Fife KY16 9SS Scotland, UK*

**Abstract:** Expanding our knowledge of student difficulties in advanced undergraduate electromagnetism is essential if we are to develop effective instructional interventions. Drawing on an analysis of course materials, in-class observations and responses to conceptual questions, we document specific resources employed by students when reasoning about the divergence of a vector field. One common student error, which persisted in our course despite explicit instruction, is to misinterpret any "spreading out" of field lines in a diagram as representing a place of non-zero divergence. Some of these student difficulties can likely be attributed to having first learned about the divergence in a mathematical context, where there was little emphasis on graphical representations of vector fields and connections to physical situations.

## INTRODUCTION

An important first step in developing effective instructional strategies and materials is to document persistent student difficulties, and how they manifest in particular topics, such as electromagnetism (E&M). Prior studies have investigated some of the more challenging concepts in advanced undergraduate E&M [1, 2], as well as students' facility with relevant mathematical tools [3, 4]. One article highlights problems that graduate students have with interpreting diagrams of vector fields with a non-zero divergence or curl, though this was primarily used to motivate a broader discussion about conceptual learning goals in graduate education [5].

We discuss here some of the ways that students think about the divergence of a vector field in advanced undergraduate E&M, at the University of St Andrews and at other institutions. We have conducted classroom observations; analyzed instructional materials; and gathered data on student responses to concept tests, homework/exam questions and a research-validated conceptual assessment [6].

Having learned about the divergence in a purely mathematical context, most of our students initially had a poor understanding of how to physically interpret Gauss' law in differential form; and had difficulty making sense of vector plots. What they did learn about E- & B-fields ($\nabla \cdot \mathbf{E} = \rho / \varepsilon_0$; $\nabla \cdot \mathbf{B} = 0$) often did not transfer to their understanding of the continuity equation ($\nabla \cdot \mathbf{J} = -\partial \rho / \partial t$). The most common and persistent student error was an automatic association of positive divergence with locations in a field diagram where lines or vectors "spread apart" from each other. For many students, such incorrect ideas about the divergence persisted despite explicit instruction.

## BACKGROUND AND INSTRUCTION

The E&M course at St Andrews covers a selection of topics at the level of Griffiths [7], in 30 lectures over an 11-week semester. A typical 50-minute class period consisted primarily of lecturing, punctuated by concept tests and student-initiated questions. Weekly problem sets were discussed in bi-weekly 1-hour recitations (~5 students) led by faculty members. The instructor (CB) and a student researcher (CA) also held twice-weekly optional homework help sessions, irregularly attended by about 20% of the class. The student researcher observed and took field notes during lectures and help sessions, and regularly consulted with another recitation instructor regarding student difficulties.

Almost all of the 85 enrolled students had studied introductory E&M in the previous academic year, covering standard topics up through Maxwell's equations in integral form. 20% were math/physics double majors who had taken a vector calculus course from the math department in the year prior; all but a few of the remaining students had taken a "Math for Physicists" (MfP) course in the previous semester, which included topics from vector calculus.

The first E&M homework was a review assignment that included questions about the divergence and curl of a vector field. Student responses showed they had little difficulty with calculating the divergence from a mathematical expression; more challenging was a question asking them to examine four diagrams and determine whether the field **F** had non-zero divergence somewhere in space [Fig. 1]. Some student difficulties stemmed from not being familiar with vector plots, and how they differ from field line diagrams. Fig. 1(a) agrees with what most students think a field with non-zero divergence ought to look like; it was useful for them to compare this with Fig. 1(b), and to draw connections between that diagram and the mathematical expression for $\nabla \cdot \mathbf{F}$ in rectangular coordinates: $\partial F_x/\partial x + \partial F_y/\partial y + \partial F_z/\partial z$.

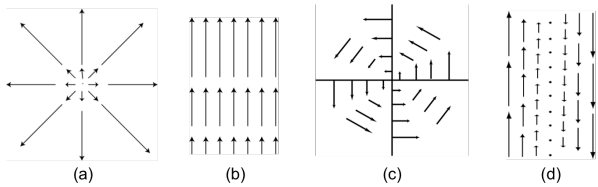

**FIGURE 1.** Diagrams from a review problem asking whether the given vector field **F** has non-zero divergence somewhere in space. For (a) & (b), $\nabla \cdot \mathbf{F} \neq 0$ everywhere; for (c) & (d), $\nabla \cdot \mathbf{F} = 0$ everywhere. [Adapted from Ref. 9.]

This task was intended to help students develop intuitions about vector fields, but there are two drawbacks to the examples chosen for this problem: (i) Fig. 1(a) likely reinforces the association of spreading field lines with positive divergence; and (ii) the combination of vector fields shown in (a) & (b) [in which $\nabla \cdot \mathbf{F} \neq 0$ everywhere] with those in (c) & (d) [in which $\nabla \cdot \mathbf{F} = 0$ everywhere] may inadvertently bolster the false impression that vector fields either have non-zero divergence or they don't, as though this were a global property of the field. This common misunderstanding has also been observed by others [8].

An examination of the lecture notes and homework from the MfP course revealed similar issues. The two examples used when introducing divergence and curl were $\mathbf{E}_1 = E_0 \hat{r}$ and $\mathbf{E}_2 = E_0 \hat{\theta}$; likewise, every homework problem involved a vector field having non-zero divergence and zero curl everywhere in space, or vice-versa. The only exception was a later assignment that involved a $1/r^2$ radial field, which has zero divergence everywhere except at the origin (the problem statement did not refer to this as the field of a point charge). The intent was to highlight an apparent inconsistency between a direct calculation of the flux and the result found using the divergence theorem. In the posted solutions, the instructor merely stated: "It turns out that […] there is a delta function 'hiding' at the origin, which resolves the apparent paradox." It should be noted that this course contained no discussion of field sources and their relationship to the divergence, and there was very little physical context provided in the section on vector calculus.

After showing that $\nabla \cdot \mathbf{E}_1 = E_0/r$ in his lecture notes, the instructor commented: "…this vector field has a non-zero divergence everywhere, although the divergence decreases as $r$ increases: this reflects the fact that 'neighboring' arrows [in the figure] are becoming more nearly parallel (non-divergent) as we get farther from the origin." This statement is not incorrect, but he failed to mention how this resource (parallel field lines → zero divergence, or its converse) can often be misleading (e.g., Fig. 1(b); see below).

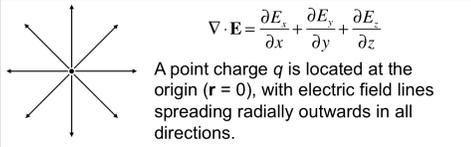

**FIGURE 2.** In-class concept test on the divergence of the field due to a point charge. The distribution of responses was: (A) 18%, (B) 0%, (C) 77%, (D) 5% [N = 82].

During the fourth E&M lecture, after students had completed the review assignment, a concept test was used to elicit and confront these issues [Fig. 2]. The expression for divergence in rectangular coordinates was deliberately placed at the top of the slide so as to activate a resource that had been useful in analyzing Fig. 1(b): examining whether the magnitude of a field component changes along that same direction. They had seen the differential form of Gauss' law several times in recent lectures, but only 18% of students answered correctly (the divergence is only non-zero at the location of a field source). The instructor encouraged students then, and throughout the semester, to take the equals sign in $\nabla \cdot \mathbf{E} = \rho/\varepsilon_0$ literally: the left- and right-hand sides have the same values at all points in space. This is true of any equation, but was not immediately obvious to many of them in this case. During the follow-up discussion in class (and in later recitations), students expressed confusion about the divergence of this field being zero along (for example) the $x$-axis; some argued that there is only a single non-zero component there, which is changing with distance, making $\partial E_x/\partial x$ non-zero. They were not recognizing that they had been implicitly evaluating the $y$- and $z$- components on the $x$-axis before taking derivatives.

As a final example related to the assessment questions discussed below, a later homework problem concerned a steady current flowing through a uniform cylindrical resistor. Students were to decide in turn which of these quantities are *zero* or *non-zero*: the time-derivative of the volume charge density $\rho$ [zero, because it is a steady state situation]; the divergence of the current density **J** [zero, by $\nabla \cdot \mathbf{J} = -\partial \rho/\partial t$]; the divergence of **E** [zero, using Ohm's law $\mathbf{J} = \sigma \mathbf{E}$]; and $\rho$ inside the resistor [zero, using $\nabla \cdot \mathbf{E} = \rho/\varepsilon_0$]. A subset of students' solutions (~25%) showed that most were able to generate the correct answers.

## ASSESSMENT RESULTS

On the midterm exam, students were asked to state where the divergence and curl of an electric dipole field are zero or non-zero, and to briefly explain their reasoning. A dipole field was chosen because the answer is straightforward using Maxwell's equations, but there is the distracting appearance of "spreading" and "curling" field lines [Fig. 3]. 59% of students received full credit, and most of the rest gained some partial credit, primarily for correct responses with incomplete reasoning; 13% received zero points. Only a handful of students attempted to reason in terms of non-zero derivatives or spreading field lines, which was taken at the time to indicate that explicit instruction had been reasonably effective.

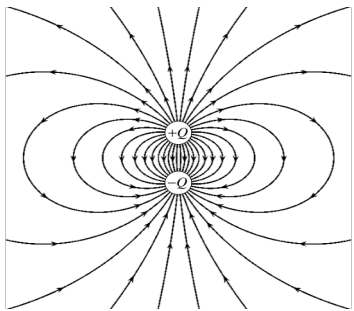

**FIGURE 3**. Diagram from a midterm exam question on the divergence and curl of an electric dipole field. $\nabla \cdot \mathbf{E} \neq 0$ only at the location of the point charges; $\nabla \times \mathbf{E} = 0$ everywhere.

During the last week of the course we administered the Colorado UppeR-division ElectromagNetism Test (CURrENT) [8]. Question 4(b) of this conceptual assessment involves a steady current in a section of wire where the diameter is gradually decreasing; it asks: "Inside this section of wire, is the divergence of the current density $\nabla \cdot \mathbf{J}$ zero or non-zero?" [Fig. 4]. Out of 63 students, only 38% responded correctly that $\nabla \cdot \mathbf{J} = 0$ inside the wire, of which 88% provided correct reasoning. The following are typical examples of *incorrect* responses:

(i) "$\nabla \cdot \mathbf{J}$ *non-zero inside because the current density converges toward the right, so at least one of its x,y,z-derivatives is non-zero.*"

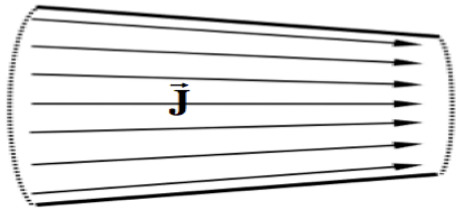

**FIGURE 4**. Diagram from Q4 of the CURrENT, showing a steady current flowing in a wire whose radius is decreasing.

(ii) "*The decreasing diameter causes the field lines to come closer together, therefore a negative divergence.*"
(iii) "*There are moving charges which are a source of divergence.*"
(iv) "*There is a net flux through a box.*"
(v) "*Divergence is a measure of how much is flowing from/into a point. In the above diagram, the lines are coming closer together. If I were to continue them, it would appear as if they were originating from a point.*" [See Fig. 5 for a student-generated diagram illustrating this line of reasoning.]

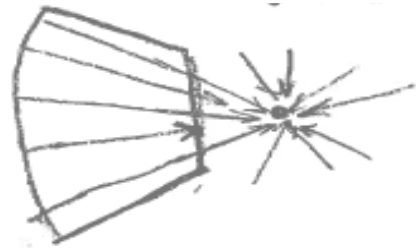

**FIGURE 5**. Drawing by a student to illustrate their reasons for concluding $\nabla \cdot \mathbf{J} \neq 0$ in Q4(b) of the CURrENT.

Such results are not specific to St Andrews; in fact, they are typical of most institutions where the CURrENT has been administered. Fig. 6 shows the percentage of students who correctly indicated that $\nabla \cdot \mathbf{J} = 0$ in Q4(b) for six different E&M classes where interactive engagement methods were used. [Course A is St Andrews; Courses B-F took place at institutions in the United States.] Courses A-C were taught by instructors with backgrounds in PER, but only Course C used in-class tutorials developed by the University of Colorado Boulder, some of which are directly relevant to student understanding of divergence and curl [9].

We also categorized the combined responses provided by those students from Courses A-C who did *not* indicate that $\nabla \cdot \mathbf{J} = 0$ inside the wire (N = 79, representing 52% of 153 students), as summarized in Table 1. Each of the illustrative student responses quoted above [(i) – (v)] corresponds to one of the first five categories listed in the table, as indicated.

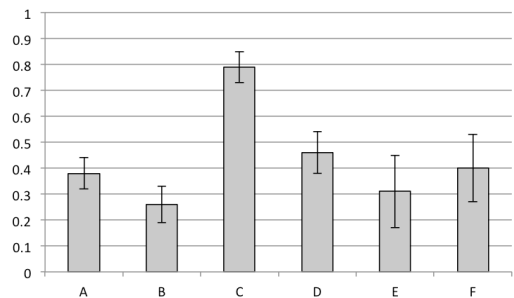

**FIGURE 6**. Percentage of students correctly stating that $\nabla \cdot \mathbf{J} = 0$ in Q4(b) of the CURrENT; error bars represent the standard error on the mean.

**TABLE 1.** Categorization of responses from students in Courses A-C who did not answer Q4(b) correctly [N = 79]. Roman numerals (i)-(v) correspond to the illustrative student responses quoted in the text.

| TYPE OF REASONING | % |
|---|---|
| Magnitude of **J** increasing to the right (i) | 47 |
| Field lines becoming more dense (ii) | 13 |
| Charge density as source of divergence (iii) | 8 |
| Non-zero net flux of **J** through a surface (iv) | 6 |
| Source/sink exists somewhere (v) | 5 |
| Other | 9 |
| Blank / no reasoning provided | 12 |

During the development of the CURrENT, validation interviews were conducted with 7 students to establish that they were interpreting these questions as intended, and also that their written responses were consistent with the reasoning they expressed verbally. This process was recently repeated with 4 students using the latest version (V.5), with the same results [6]. We would therefore argue that the categorization of written responses in Table 1 is an adequate reflection of student thinking regarding the divergence in this specific context. Throughout its history, this question has remained essentially unchanged, so we were also able to determine the percentage of correct responses for a larger and more diverse population. Across eight different institutions, only 146 from a total of 376 students (=39%) correctly stated that $\nabla \cdot \mathbf{J} = 0$ in this situation.

## DISCUSSION AND CONCLUSIONS

The data presented here show that students may rely on a number of ideas when reasoning about the divergence of a vector field, some of which are productive in specific contexts, but not universally. Most students at St Andrews recognized during the midterm exam that the divergence of a field is only non-zero at the location of a source (when asked about an electric dipole field), but many did not access this same resource when later considering the current density inside a wire, even though they had all seen a similar homework problem (albeit one that did not contain the distraction of "converging" field lines). Instead, they appealed to other resources that are not necessarily wrong in and of themselves, but lead to erroneous conclusions when applied in the wrong situations. In other words, there were many students who understood the meaning of divergence in the context of Gauss' law, but in other contexts did not employ an epistemological framing [10] that led them to access the relevant knowledge (e.g., not recognizing the significance of the word "steady" in the problem statement of Q4).

Despite instruction to the contrary, a significant number of students expressed incorrect reasoning, primarily having to do with the divergence being non-zero when at least one of the components of a field is changing with distance (non-zero partial derivatives), or wherever field lines in a diagram are becoming more or less closely spaced. Students were easily distracted by the *semblance* of a divergence in a field; and some believed the existence of a source/sink at one point in space implies the divergence is everywhere non-zero. Our experience suggests that students face analogous difficulties with the curl of a vector field, which requires further investigation.

Although we have insufficient evidence to demonstrate a direct link between specific instructional choices and the student thinking described herein, there are good reasons to believe that when students are introduced to the concept of divergence without reference to field sources, and without explicit attention paid to common difficulties, they can develop robust misconceptions that will be resistant to later instruction. We suggest that instructors be conscious of this when developing the tools of vector calculus in the physical context of electromagnetism; and should explicitly address these specific student difficulties within a variety of topics, in order for students to develop proficiency in determining when a particular conceptual resource will be productive.

## ACKNOWLEDGMENTS

Many thanks to C. Hooley for useful discussions, and to S. Pollock and Q. Ryan for feedback and assessment scoring. We greatly appreciate the cooperation of the students and instructors who made this study possible.

## REFERENCES

1. C. Wallace & S. Chasteen, *PRST-PER* **6**, 020115 (2010)
2. C. Manogue, et al., *Am. J. Phys.* **74**, 344 (2006).
3. R. Pepper, et al., *PRST-PER* **8**, 010111 (2012)
4. B. Wilcox, et al., *PRST-PER* **9**, 020119 (2013).
5. C. Singh and A. Maries, *PERC Proc. 2012* (AIP, Melville NY, 2013), p. 382.
6. Q. Ryan, et al., *PERC Proc. 2014* (accepted); see also http://per.colorado.edu/Electrodynamics/.
7. D. Griffiths, *Introduction to Electrodynamics, 3rd Ed.* (Prentice-Hall, Upper-Saddle River NJ, 1999).
8. www.physics.oregonstate.edu/portfolioswiki/activities: reflections:local:vfdivergence
9. per.colorado.edu/Electrodynamics/tutorials.html
10. E. F. Redish, *Am. J. Phys.* **82**, 537 (2014).